\documentclass{iau_FM}
\usepackage{graphicx}

\title[Small-scale clustering of nano-dust grains]{Small-scale clustering of nano-dust grains in turbulent interstellar molecular clouds
[\textit{Extended version}]}

\author[Lars Mattsson] 
{Lars Mattsson}

\pubyear{2018}
\affiliation{Nordita, KTH Royal Institute of Technology \& Stockholm University, Sweden \\email: {\tt lars.mattsson@nordita.org}} 
\begin{document}

\maketitle

\begin{abstract}
Clustering and dynamics of nano-sized particles (nano dust) is investigated using high-resolution ($1024^3$) simulations of compressible isothermal hydrodynamic turbulence, intended to mimic the conditions inside cold molecular clouds in the interstellar medium. Nano-sized grains may cluster in a turbulent flow (small-scale clustering), which increases the local grain density significantly. Together with the increased interaction rate due to turbulent motions, aggregation of interstellar nano-dust may be plausible. 
\end{abstract}

\section{Introduction}
Nano-sized dust grains, like graphitic particles, polycyclic aromatic hydrocarbons (PAHs) or even nano diamonds, are by number the most abundant type of dust in the interstellar medium (ISM). These grains make up only a tiny fraction of the total interstellar dust mass, but are believed to be a significant source of extinction. This would certainly be the case if the grains are uniformly and isotropically distributed in the ISM, but if they are concentrated into small clumps and filaments with essentially dust free regions in between, the net extinction of incident radiation may be significantly reduced. Moreover, increased grain densities means the grain-grain interaction rate will also increase, in particular in the inner regions of cold molecular clouds (MCs). How nano dust will cluster (or not) is therefore important to determine. 

It is well established that large grains will decouple from a turbulent gas flow, while small grains will tend to trace the motion of the gas as seen in, e.g., the recent simulations by \cite{Hopkins16} and \cite{Mattsson18}. Small grains may still cluster on scales smaller than those typical for a turbulent flow due to centrifuging of particles away from vortex cores and accumulation of particles in convergence zones (a.k.a. preferential concentration). Simulations of incompressible turbulence have shown that inertial particles, like interstellar dust grains, will cluster in fractal patterns with (correlation) dimensions significantly less than  the geometric dimension (see, e.g., \cite{Bhatnagar16}). However, this has not yet been demonstrated explicitly for \textit{compressible} flows. 

The present paper presents three-dimensional (3D)  high-resolution ($1024^3$) direct numerical simulations of clustering and dynamics of nano dust embedded in a turbulent ideal non-magnetic and neutral gas, with an average Mach number $\mathcal{M}_{\rm rms} = 3.2$. 

\section{Modelling and method}
The clustering and dynamics of nano-dust, implemented as inertial particles, are studied in forced homogeneous isothermal turbulence (mimicking the conditions in centres of MCs) by simulations made with the PENCIL code. This is a non-conservative, high-order, finite-difference code (sixth order in space and third order in time) described in \cite{Brandenburg02}. For more information on the type of turbulence simulations used in this study, see also \cite{Mattsson18}.

Inertial particles suspended in a gaseous medium will show a delayed response to kinetic drag from gas particles. The simplest equation of motion for dust particles is therefore
\begin{equation}
\label{stokeseq}
{d \mathbf{v}\over d t}  = {\mathbf{u}-\mathbf{v}\over \tau_{\rm s}},
\end{equation}
where $\mathbf{v}$ and $\mathbf{u}$ are the velocities of the dust and the gas, respectively, and $\tau_{\rm s}$ is the stopping time, i.e., the timescale of acceleration (or deceleration) of the grains. In the simulations $\tau_{\rm s}$ is a function of the ``grain-size parameter'', 
\begin{equation}
\alpha = {\rho_{\rm gr}\over\langle \rho\rangle}{a\over L} ,
\end{equation}
where $a$ is the radius, $\rho_{\rm gr}$ is the material density of the grains and $L$ is the length of the side of the simulation box. Here, it is noteworthy that $a$ and $\rho_{\rm gr}$ basically determine the value of $\tau_{\rm s}$ in a simulation, since $\tau_{\rm s} \sim \alpha$, to first order. However, the stopping time depends not only on the size and density of the grain, but also on the gas density and the relative Mach number $\mathcal{W}_{\rm s} = |\mathbf{u}-\mathbf{v}|/c_{\rm s}$, where $c_{\rm s}$ is the sound speed \cite{Schaaf63}. For very small grains $\mathcal{W}_{\rm s} \ll 1$ on average, but local and temporary effects may still play a role. Thus, two kinds of prescription formulae for kinetic Epstein-type drag in three cases are considered:\\
\begin{itemize}
\item {\bf Case I:}~Stopping time prescription for $\mathcal{W}_{\rm s} \ll 1$ and turbulence induced by purely compressive forcing.
\item {\bf Case II:} ~Stopping time prescription including correction for large $\mathcal{W}_{\rm s}$ and turbulence induced by purely compressive forcing.
\item {\bf Case IIb:}~Same as above, but solenoidal instead of compressive forcing. 
\end{itemize}
~

The degree of clustering is quantified by determining the correlation dimension $D_2$ of the 3D dust distribution (Mattsson et al. 2018b). $D_2$ is a kind of fractal dimension and as such it does not measure compaction of the grains due to gas compression; it only measures the fractal clustering caused by turbulence. For more details about how $D_2$ is determined, see Mattsson et al. (2018a,b).

\begin{figure}[h]
\vspace*{-0.2 cm}
\begin{center}
 \includegraphics[width=6.67cm]{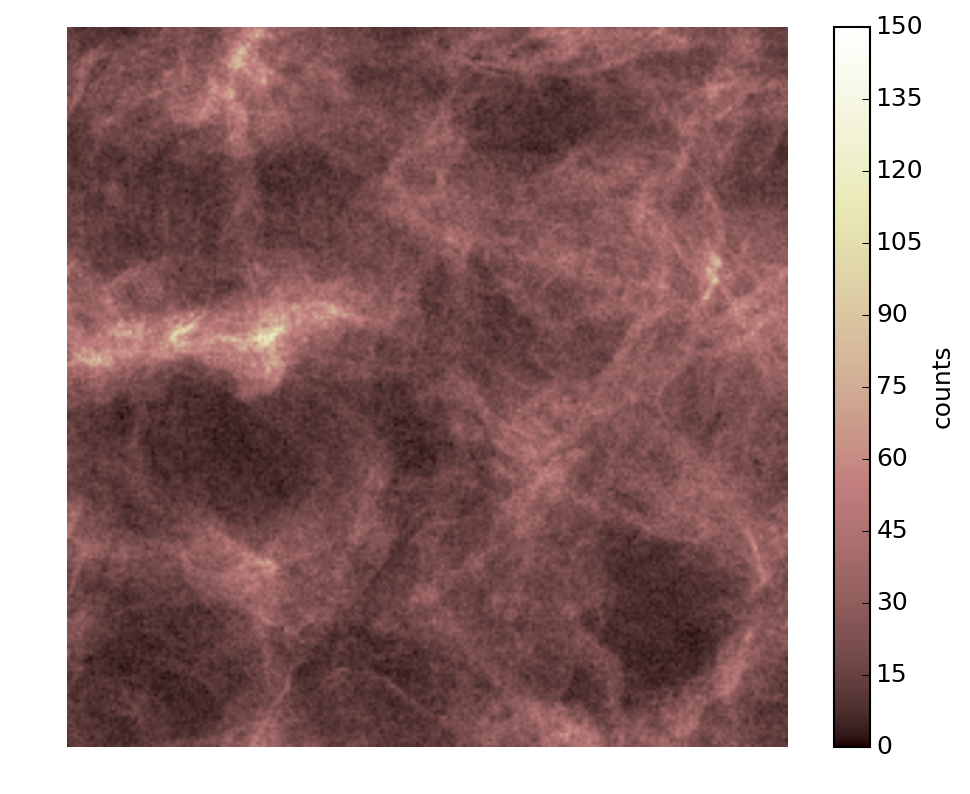} 
  \includegraphics[width=6.67cm]{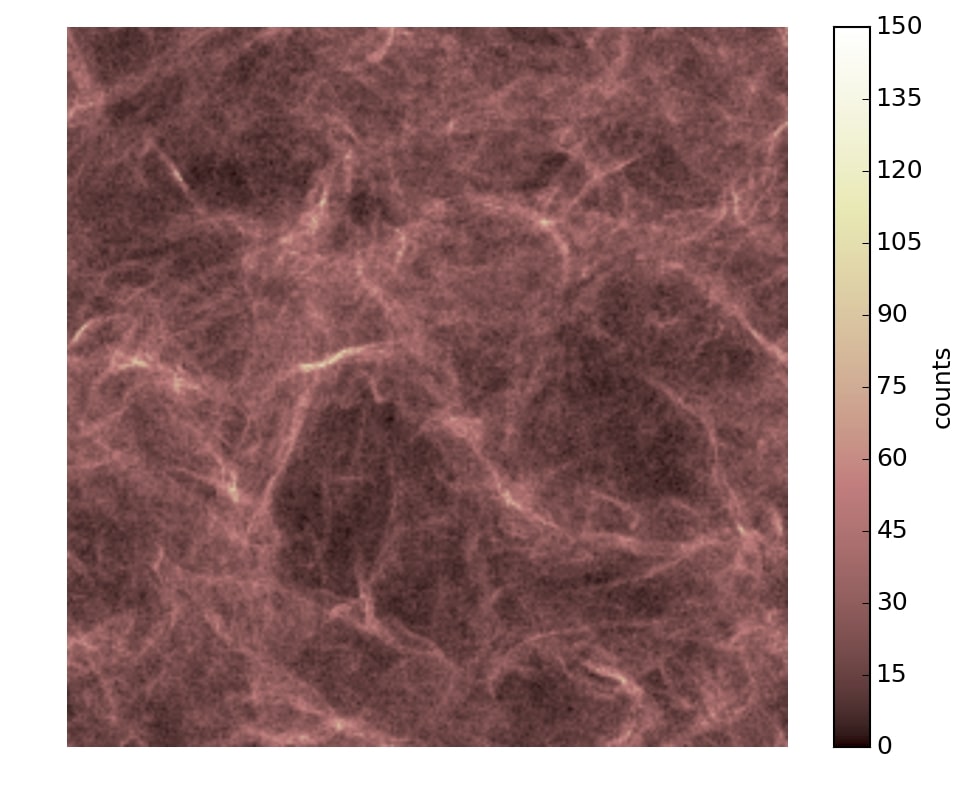} 
 \caption{Projected number density of the maximally clustered dust grains ($\alpha = 0.016$, $a \approx 12$~nm and $\alpha = 0.033$, $a \approx 25$~nm, respectively) in simulations using the transsonic approximation for the kinetic drag. Left panel shows the run with compressive forcing (Case II), while the right panel shows the run with solenoidal forcing (Case IIb).}
   \label{fig1}
\end{center}
\end{figure}

\section{Results and discussion}
Clear fractal clustering can be seen in all three cases; examples are shown in Fig. \ref{fig1} for Case II and IIb. The case with solenoidal forcing (Case IIb) appears to yield a somewhat more ``threaded'' distribution of grains, with more fine structure. This is likely the result of a higher degree of vorticity, which should result in more particles accumulating in the convergence zones in between vortices \cite{Eaton94}. 

\begin{figure}[h]
\vspace*{-0.2 cm}
\begin{center}
 \includegraphics[width=6.67cm]{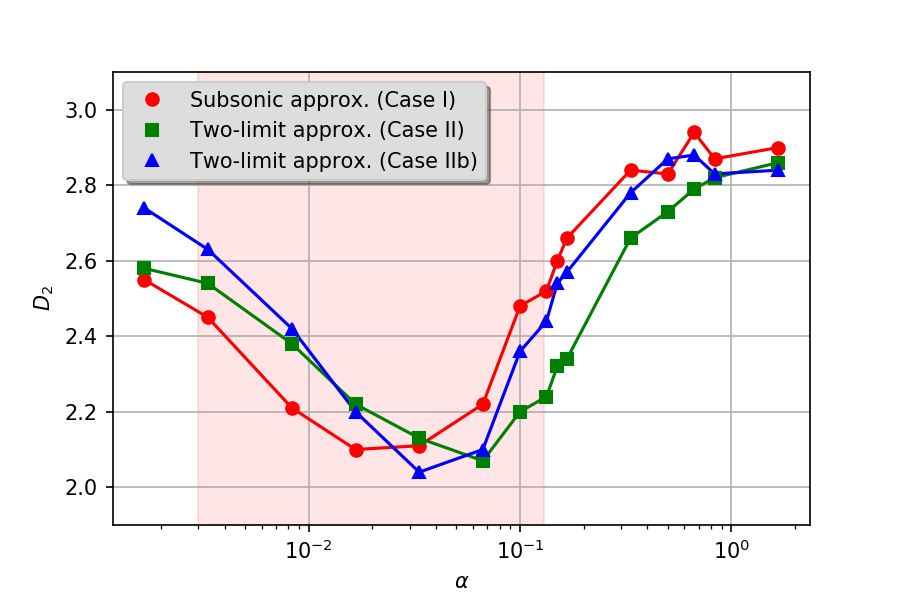} 
  \includegraphics[width=6.67cm]{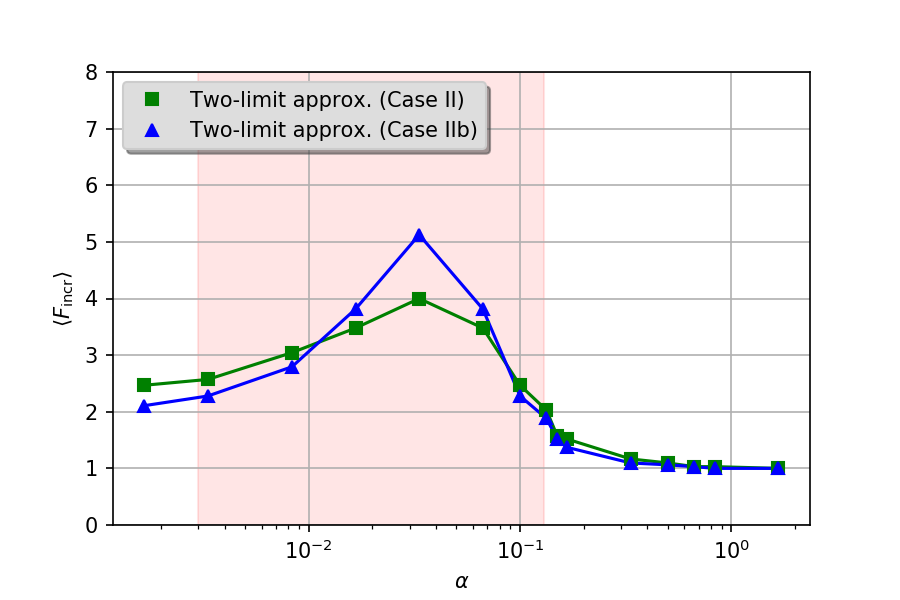} 
 \vspace*{-0.1 cm}
 \caption{Correlation dimension $D_2$ (left) and the average relative increase of the dust density $\langle F_{\rm incr} \rangle$ (right) as a function of the grain-size parameter $\alpha$ for three cases: compressive forcing and a subsonic approximation for the kinetic drag (Case I); a transsonic approximation for the drag (Case II); solenoidal forcing and the transsonic approximation (Case IIb).}
   \label{fig2}
\end{center}
\end{figure}

The maximal degree of clustering (or minimal $D_2$) is basically the same in all three simulations, but $D_2$ as a function of grain size ($\alpha$) differs slightly in shape. Fig. \ref{fig2} shows the correlation dimension $D_2$ and the average relative increase of the dust density $\langle F_{\rm incr} \rangle$ as a function of $\alpha$ for simulations with different assumptions regarding kinetic drag and forcing of the flow. The clustering has a maximum for around a certain $\alpha$, which lies in the nano-dust range (shaded area) for a typical mass-scaling of the simulations. Combined with the fact that nano dust may be abundant, and the increased interaction rate due to turbulent motions, the values of $\langle F_{\rm incr} \rangle$ suggest aggregation of nano dust may be quite efficient. Comparing coagulation models based on the MOMIC code by Mattsson (2016), with and without corrections for turbulent clustering and relative motion of nano dust, we can see an order-of-magnitude increase of the coagulation rate.

Additional forces due to magnetic fields  that may be induced due to turbulence have not been considered. Neutral, non-magnetic grains should of course not be much affected, but electrically charged grains and grains with magnetic dipole moments  can have qualitatively different dynamics due to Coulomb and Lorentz forces acting on the grains in addition to the kinetic drag force \cite{Draine03}. Even in a cold neutral medium, nano-dust grains will experience significant charge fluctuations. The charge distribution of grains may be clearly weighted towards neutral grains, but still those grains do not remain neutral for very long. Charge fluctuations are very fast and one can thus assume that individual dust grains always carry a net average charge. For this reason, magnetic fields will play a role under most circumstances and the simulations of clustering of nano-sized grain due to kinetic drag only, like those presented here, should be regarded as a bench mark to compare with previous work and the existing theory for clustering in incompressible turbulence.

In addition to charge fluctuations, it is also well known that very small dust grains undergo significant temperature fluctuations \cite{Draine03}. All populations of dust grains have a grain temperature distribution (GTD) due to differences in grain sizes. For larger grains the size dispersion essentially determines the GTD \cite{Mattsson15}, while for nano-sized grains, even a mono-dispersed population would show a wide GTD due to temperature fluctuations. Clustered nano dust will cause limited extinction, in the same way as when the dust mass is concentrated in large grains, while still emitting significant infrared emission. The GTD of nano-dust grains could make them mimic the infrared flux of a multidispersed population of larger grains, but the total mass is of course much less for any realistic number density. There may therefore be an observational degeneracy between dust emission due to tiny, strongly clustered grains and larger, more  randomly distributed grains, which raises questions regarding the connection between infrared flux and dust masses beyond the basic analysis by Mattsson et al. (2015).

\section{Summary and conclusions}
Simulations of kinetic drag on nano-sized particles (dust grains) in compressible turbulence, show maximal clustering for smaller dust particles compared to clustering of similar inertial particles in simulations of incompressible turbulence. Nano-dust grains are inertial particles despite their tiny masses, which is illustrated by the fact that the correlation dimension $D_2 <3$ over the whole nano-dust range (i.e., grains with radii in the range $a = 1\dots100$~nm) assuming a scaling of the simulations which represents the conditions in the central regions of molecular clouds. 

The average compression/compaction factor $\langle F_{\rm incr} \rangle$ for dust reaches a maximum for grain sizes near the minimum of $D_2$ (maximal clustering). Grains with radii of a few hundreds of of a micron ($a \sim 20-60$~nm) show  a number density increase of a factor 4--5. Thus,  the probability that grains in this size range will interact with another grain increases by a similar factor. As a direct consequence, the expected rate of coagulation increases enough for aggregate formation by nano dust in MCs to be regarded as feasible, even if the target grain is initially quite small. 

Charged nano-dust grains will have a  different behaviour compared to the passive-scalar type dust in the present simulations. Whether this will further accentuate the clustering or lead to dispersion, counteracting the small-scale clustering, is unclear. Determining this will be the goal of future simulations.

\end{document}